\documentclass[preprint,showpacs,showkeys,nobibnotes,nofootinbib,eqsecnum,floats,onecolumn,unsortedaddress,byrevtex,aps]{revtex4}

\usepackage{amsfonts}

\input{tcilatex}

\begin{document}

\preprint{ENSK--TP--41}
\title{MONTE CARLO STUDY OF FOUR-SPINON DYNAMIC STRUCTURE FUNCTION IN
ANTIFERROMAGNETIC HEISENBERG MODEL}
\author{A. Abada}
\email{abada@wissal.dz}
\affiliation{D\'{e}partement de Physique, Ecole Normale Sup\'{e}rieure, BP 92
Vieux-Kouba, Alger 16050, Algeria}
\author{B. Si-Lakhal}
\email{silakhal@wissal.dz}
\affiliation{D\'{e}partement de Physique, Universit\'{e} de Blida,
BP 270 Blida 09000, Algeria}
\date{\today}

\begin{abstract}
Using Monte Carlo integration methods, we describe the behavior of the exact
four-spinon dynamic structure function $S_{4}$ in the antiferromagnetic spin
1/2 Heisenberg quantum spin chain as a function of the neutron energy
$\omega$ and momentum transfer $k$. We also determine the four-spinon
continuum, the extent of the region in the $\left( k,\omega \right)$ plane
outside which $S_{4}$ is identically zero. In each case, the behavior of
$S_{4}$ is shown to be consistent with the four-spinon continuum and compared
to the one of the exact two-spinon dynamic structure function $S_{2}$.
Overall shape similarity is noted.
\end{abstract}

\maketitle

\section{Introduction}

The spin $s=\frac{1}{2}$ Heisenberg quantum spin chain describes the
magnetic properties of quasi-one-dimensional antiferromagnetic compounds
like KCuF$_{3}$ \cite{hirakawa-kurogi}. The spin dynamics is experimentally
investigated using inelastic neutron scattering \cite{Exp}. From a
theoretical standpoint, the quantity of interest is the dynamic structure
function (DSF) $S$ of two local spin operators. This is because the
magnetic scattering cross section per magnetic site is directly proportional
to it \cite{squires-lovesey}.

The Heisenberg model has been studied quite intensively \cite{PSD,Baxter},
and because of the presence of the
$\mathrm{U}_{q}(\widehat{\mathrm{sl}}_{2})$ symmetry \cite{JimMi},
a number of exact results are available. Static properties need only
the Yang-Baxter relation \cite{Baxter}, whereas dynamic correlation functions
require the additional notion of vertex operators and exploit bosonization
methods \cite{CorFun}. One thus obtains
compact expressions for form factors \cite{JimMi}.

Regarding the dynamic structure function, the focus has so far been on
$S_{2}$, the two-spinon contribution to the total $S$. First there has been
the Anderson (semi-classical) spin-wave theory \cite{And}, an approach based
on an expansion in powers of $1/s$ and hence, exact only in the classical
limit $s=\infty $. It can describe with some satisfaction compounds with
higher spins \cite{TMMC}, but fails in the quantum limit $s=\frac{1}{2}$.
For this latter system, the M\"{u}ller ansatz has been proposed, built
mainly from finite-chain calculations and symmetry considerations
\cite{Muller}. It is an approximate expression for $S_{2}$ that accounts
satisfactorily for many aspects of the phenomenology \cite{Exp}. More
recently, an exact expression for $S_{2}$ has been obtained, making
extensive use of the $\mathrm{U}_{q}(\widehat{\mathrm{sl}}_{2})$ symmetry 
\cite{BCK}, and a comparison with the M\"{u}ller ansatz shows that it gives
a better account of the data \cite{KMB,BKM}.

Beyond the two-spinon DSF $S_{2}$ are of course all the other $S_{2p>2}$
contributions. The first one to look at is the four-spinon dynamic structure
function $S_{4}$, and the purpose of the present article is to describe its
behavior. An exact expression for $S_{4}$ has been derived in \cite{ABS},
see also \cite{Boug}. We intend to use that expression to describe the
behavior of $S_{4}$ as a function of the neutron energy transfer $\omega$
and neutron momentum transfer $k$.

We must mention that a preliminary investigation into the behavior of $S_{4}$
has already been initiated in \cite{ABSS}. There we have used quadratures to
perform the integrations involved in the expression of $S_{4}$. But because
of slow convergence of the algorithms we wrote, we could describe the
behavior of $S_{4}$ only as a function of $k$, and only for relatively small
values of $\omega$. In the present work, the integrations are performed
using Monte Carlo methods, and this makes it possible not only to study
$S_{4}$ as a function of $k$ for a wider range of values of $\omega$, but
obtain a description of $S_{4}$ as a function of $\omega$ for a wide range
of values of $k$ as well. It is however important to note that, for the same
values of $\omega$, the behavior of $S_{4}$ as a function of $k$ we obtain
here by Monte Carlo methods is similar and consistent with the one we
obtained in \cite{ABSS} using quadratures.

Also, we systematically carry a comparison of each result we obtain
regarding $S_{4}$ to a corresponding one regarding $S_{2}$. The reason is
that there is good familiarity in the literature with the two-spinon DSF and
so, such a comparison allows faster acquaintance with the four-spinon
contribution. To have the discussion as clear as possible, we scale both
$S_{4}$ and $S_{2}$ to one. All results concerning $S_{2}$ are already
known \cite{BCK,KMB}; only those concerning $S_{4}$ are new.

This article is organized as follows. After these introductory remarks, we
describe in the next section the Heisenberg model and give the definition of
the dynamic structure function. We write its decomposition in $2p$-spinon
contributions and give the expressions of $S_{2}$ and $S_{4}$. In section
three, we first determine the four-spinon continuum, the region in the
$\left( k,\omega \right)$ plane outside which $S_{4}$ is identically zero.
We will see that it extends beyond the spin-wave des Cloizeaux and Pearson
boundaries. To the best of our knowledge, this result is a first direct and
explicit exact theoretical confirmation that the total $S$ for the infinite
chain tails outside the spin-wave continuum. Next in this section is a
description of the behavior of $S_{4}$ as a function of $\omega$ for fixed
values of $k$ followed by a comparison with the corresponding behavior of
$S_{2}$. It is seen that there is consistency with the four-spinon continuum
and that the overall shape (not the details) of $S_{4}$ is similar to the
one of $S_{2}$, each in its own respective continuum. Last is carried a
description of the behavior of $S_{4}$ as a function of $k$ for fixed values
of $\omega$ and a comparison with the corresponding one of $S_{2}$. Here
too similarity between the two overall shapes is found as well as
consistency with the four-spinon continuum. Section four includes concluding
remarks and indicates few directions in which this work can be carried
forward.

This paper is a continuation of the work \cite{ABS}. We use the same
notation except for a slight modification in (\ref{FS4}) where here we
introduce the function $h$ instead of the function $f$ with the relation
$f\equiv \exp \left( -h\right)$.

\section{Four-spinon dynamic structure function}

The antiferromagnetic spin-$\frac{1}{2}$ $XXX$ Heisenberg chain is defined
as the isotropic limit of the $XXZ$ anisotropic Heisenberg Hamiltonian: 
\begin{equation}
H=-\frac{1}{2}\hspace{-3pt}\sum_{n=-\infty }^{\infty }\left( \sigma
_{n}^{x}\sigma _{n+1}^{x}\hspace{-3pt}+\sigma _{n}^{y}\sigma _{n+1}^{y}
\hspace{-3pt}+\Delta \sigma _{n}^{z}\sigma _{n+1}^{z}\right) \,.
\label{hamiltonian}
\end{equation}
$\Delta =(q+q^{-1})/2$ is the anisotropy parameter and the isotropic
antiferromagnetic limit is obtained as $\Delta \rightarrow -1^{-}$, or
equivalently $q\rightarrow -1^{-}$. Here $\sigma _{n}^{x,y,z}$ are the usual
Pauli matrices acting at the site $n$ of the chain. The exact
diagonalization of this Hamiltonian is performed directly in the
thermodynamic limit. This is necessary if we want to exploit the
$\mathrm{U}_{q}(\widehat{\mathrm{sl}}_{2})$ quantum group symmetry present
in the model \cite{JimMi}. One consequence is the appearance of two vacuum
states $|0\rangle _{i}$, $i=0,1$ due to two different boundary conditions
on the infinite chain. The Hilbert space $\mathcal{F}$ consists of $n$-spinon
energy eigenstates $|\xi _{1},...,\xi _{n}\rangle _{\epsilon
_{1},...,\epsilon _{n};\,i}$ such that: 
\begin{equation}
H|\xi _{1},...,\xi _{n}\rangle _{\epsilon _{1},...,\epsilon
_{n};\,i}=\sum_{j=1}^{n}e(\xi _{j})|\xi _{1},...,\xi _{n}\rangle _{\epsilon
_{1},...,\epsilon _{n};\,i\;},  \label{energy-eigenstates}
\end{equation}
where $e(\xi _j)$ is the energy of spinon $j$ and $\xi _{j}$ is a spectral
parameter living on the unit circle. In the above relation, $\epsilon
_{j}=\pm 1$. The translation operator $T$ which shifts the spin chain by one
site acts on the energy eigenstates in the following manner: 
\begin{equation}
T|\xi _{1},...,\xi _{n}\rangle _{\epsilon _{1},...,\epsilon
_{n};\,i}=\prod_{i=1}^{n}\tau (\xi _{i})|\xi _{1},...,\xi _{n}\rangle
_{\epsilon _{1},...,\epsilon _{n};1-i\;},  \label{action-of-T}
\end{equation}
where $\tau (\xi _{j})=e^{-ip(\xi _{j})}$ and $p(\xi _{j})$ is the lattice
momentum of spinon $j$. The exact expressions of the spinon energy and
lattice momentum in terms of the spectral parameter are known in the
literature \cite{Fadeev-Takhtajan,JimMi,ABS}. We are interested in their
$XXX$ limit and it is given below in eq (\ref{dispersion-relation}). The
completeness relation in $\mathcal{F}$ reads: 
\begin{equation}
\mathbf{I}=\sum_{i=0,1}\sum_{n\geq 0}\sum_{\{\epsilon _{j}=\pm 1\}_{j=1,n}}
\frac{1}{n!}\oint \prod_{j=1}^{n}\frac{d\xi _{j}}{2\pi i\xi _{j}}\;|\xi
_{1},...,\xi _{n}\rangle _{\epsilon _{1},...,\epsilon _{n};\,i\;i;\,\epsilon
_{1},...,\epsilon _{n}}\langle \xi _{1},...,\xi _{n}|\,.
\label{completeness-relation}
\end{equation}

The two-point dynamic structure function is defined as the Fourier transform
of the zero-temperature vacuum-to-vacuum two-point function. The transverse
DSF is therefore defined by: 
\begin{equation}
S^{i,+-}(\omega ,k)=\int_{-\infty }^{\infty }dt\sum_{m\in \mathbb{Z}}
e^{i(\omega t+km)}\,_{i}\langle 0|\sigma _{m}^{+}(t)\,\sigma
_{0}^{-}(0)|0\rangle _{i}\;,  \label{definition-of-S}
\end{equation}
where $\omega$ and $k$ are the neutron energy and momentum transfer
respectively, and $\sigma ^{\pm }$ denotes $(\sigma ^{x}\pm i\sigma ^{y})/2$.
The DSF satisfies the following relations:
\begin{equation}
S^{i,+-}(\omega ,k)=S^{i,+-}(\omega ,-k)=S^{i,+-}(\omega ,k+2\pi )\,,
\label{symmetries-of-DSF}
\end{equation}
expressing reflection symmetry and periodicity. Inserting the completeness
relation (\ref{completeness-relation}) and using the Heisenberg relation: 
\begin{equation}
\sigma _{m}^{x,y,z}(t)=\exp \left( iHt\right) \,T^{-m}\sigma
_{0}^{x,y,z}(0)\,T^{m}\,\exp \left( -iHt\right) \,,
\label{heisenberg-relation}
\end{equation}
we can write the transverse DSF as the sum of $n$-spinon contributions: 
\begin{equation}
S^{i,+-}(\omega ,k)=\sum_{n\,\,\mathrm{even}}S_{n}^{i,+-}(\omega ,k)\;,
\label{sum-over-n-spinons}
\end{equation}
where the $n$-spinon DSF $S_{n}$ is given by: 
\begin{eqnarray}
S_{n}^{i,+-}(\omega ,k) &=&\frac{2\pi }{n!}\sum_{m\in \mathbb{Z}}
\,\sum_{\epsilon _{1},...,\epsilon _{n}}\oint \prod_{j=1}^{n}\frac{d\xi _{j}}
{2\pi i\xi _{j}}\,e^{im(k+\sum_{j=1}^{n}p_{j})}\,\delta \left( \omega
-\sum_{j=1}^{n}e_{j}\right) \,  \nonumber \\
&&\times \;X_{\epsilon _{n},...,\epsilon _{1}}^{i+m}(\xi _{n},...,\xi
_{1})\;X_{\epsilon _{1},...,\epsilon _{n}}^{1-i}(-q\xi _{1},...,-q\xi
_{n})\;,  \label{expression-of-Sn}
\end{eqnarray}
a relation in which $X^{i}$ denotes the form factor: 
\begin{equation}
X_{\epsilon _{1},...,\epsilon _{n}}^{i}(\xi _{1},...,\xi _{n})\equiv
\,_{i}\langle 0|\sigma _{0}^{+}\left( 0\right) |\xi _{1},...,\xi _{n}\rangle
_{\epsilon _{1},...,\epsilon _{n};\,i}\;.  \label{form-factor}
\end{equation}
In relation (\ref{expression-of-Sn}), $i+m$ is to be read modulo 2. Note
that each $S_{n}$ satisfies the symmetry relations (\ref{symmetries-of-DSF}).

An exact expression for the form factor $X^{i}$ is known \cite{smirnov,JimMi}.
To arrive at the result, one has to exploit extensively the
infinite-dimensional representation of
$\mathrm{U}_{q}(\widehat{\mathrm{sl}}_{2})$ and bosonize the relevant
vertex operators in order to be able to
manipulate in a systematic way traces of these operators which ultimately
yield the correlation functions. Using this form factor, it is possible to
give an exact expression for the $n$-spinon DSF in the anisotropic case
\cite{ABS} and determine its isotropic limit \cite{Boug}, obtained via the
replacement \cite{JimMi,ABS}: 
\begin{equation}
\xi =ie^{-2i\varepsilon \rho }\,;\qquad q=-e^{-\varepsilon }\,,\qquad
\varepsilon \rightarrow 0^{+}\,,  \label{Isotropic-limit}
\end{equation}
where $\rho$ becomes the spectral parameter suited for this limit. The
expressions of the energy $e$ and momentum $p$ in terms of $\rho$ then
read: 
\begin{equation}
e(\rho )=\frac{\pi }{\cosh (2\pi \rho )}=-\pi \sin \,p\;;\quad \cot
\,p=\sinh (2\pi \rho )\;;\quad -\pi \leq p\leq 0\;.
\label{dispersion-relation}
\end{equation}

It turns out that the transverse two-spinon DSF $S_{2}$ does not involve a
contour integration, see (\ref{expression-of-Sn}). Its exact expression has
been derived in \cite{BCK}. It reads: 
\begin{equation}
S_{2}^{+-}(\omega ,k-\pi )=\frac{1}{4}\frac{e^{-I(\rho )}}{\sqrt{\omega
_{2u}^{2}-\omega ^{2}}}\,\Theta (\omega -\omega _{2l})\,\Theta (\omega
_{2u}-\omega )\;,  \label{S2exact}
\end{equation}
where $\Theta$ is the Heaviside step function and the function $I(\rho )$
is given by: 
\begin{equation}
I(\rho )=\int_{0}^{+\infty }\frac{dt}{t}\frac{\cosh (2t)\,\cos (4\rho t)-1}
{\,\sinh (2t)\cosh (t)}\,e^{t}\,.  \label{I-de-rho}
\end{equation}
$\omega _{2u(l)}$ is the upper (lower)\ bound of the two-spinon excitation
energies called the des Cloizeaux and Pearson (dCP)\ \cite{Muller,BCK} upper
(lower) bound or limit. They read: 
\begin{equation}
\omega _{2u}=2\pi \sin \frac{k}{2};\qquad \quad \omega _{2l}=\pi \,|\sin
\,k|\,.  \label{dCP}
\end{equation}
The quantity $\rho $ is related to $\omega$ and $k$ by the relation: 
\begin{equation}
\cosh \,\pi \rho =\sqrt{\frac{\omega _{2u}^{2}-\omega _{2l}^{2}}{\omega
^{2}-\omega _{2l}^{2}}}\,,  \label{relation-rho-omega-k}
\end{equation}
which is obtained using eq (\ref{dispersion-relation}) and the
energy-momentum conservation laws. The properties of $S_{2}$ have been
discussed in \cite{KMB,BKM} where a comparison with the M\"{u}ller ansatz 
\cite{Muller} is carried.

The four-spinon DSF $S_{4}$ involves only one contour integration and its
expression is given in \cite{ABS}. For $0\leq k\leq \pi $ it reads: 
\begin{equation}
S_{4}^{+-}(\omega ,k-\pi )=C_{4}\int_{-\pi }^{0}dp_{3}\int_{-\pi
}^{0}dp_{4}\,F(\rho _{1},...,\rho _{4})\,.  \label{S4}
\end{equation}
For other values of $k$, it extends by symmetry using
(\ref{symmetries-of-DSF}). $C_{4}$ is a numerical constant irrelevant for the
present work since we will scale $S_4$ to unity, and the integrand $F$
is given by:
\begin{equation}
F(\rho _{1},...,\rho _{4})=\sum_{(p_{1},\,p_{2})}\frac{\exp [-h(\rho
_{1},...,\rho _{4})]\,\sum_{\ell =1}^{4}|g_{\ell }(\rho _{1},...,\rho
_{4})|^{2}}{\sqrt{W_{u}^{2}-W^{2}}}\,.  \label{FS4}
\end{equation}
The different quantities involved in this expression are as follows. 
\begin{eqnarray}
W &=&\omega +\pi \,(\sin \,p_{3}+\sin \,p_{4}\,)\,;  \nonumber \\
W_{u} &=&2\pi \left\vert \sin \,\left( K/2\right) \right\vert \,,\qquad
K=k+p_{3}+p_{4}\,;  \nonumber \\
\cot \,p_{j} &=&\sinh (2\pi \rho _{j})\,,\qquad -\pi \leq p_{j}\leq 0\,.
\label{arguments-of-F}
\end{eqnarray}
The function $h$ is given by: 
\begin{equation}
h(\rho _{1},...,\rho _{4})=\sum_{1\leq j<j^{^{\prime }}\leq 4}I\,(\rho
_{jj^{\prime }})\,,  \label{f}
\end{equation}
where $\rho _{jj^{\prime }}=\rho _{j}-\rho _{j^{\prime }}$ and the function
$g_{\ell }$ reads: 
\begin{eqnarray}
g_{\ell } &=&(-1)^{\ell +1}(2\pi )^{4}\sum_{j=1}^{4}\cosh \,\,(2\pi \rho
_{j})  \nonumber \\
&&\times \sum_{m=\Theta (j-\ell )}^{\infty }\frac{\prod_{i\neq \ell }(m-
\frac{1}{2}\Theta (\ell -i)+i\rho _{ji})}{\prod_{i\neq j}\pi ^{-1}\sinh (\pi
\rho _{ji})}\prod_{i=1}^{4}\frac{\Gamma (m-\frac{1}{2}+i\rho _{ji})}{\Gamma
(m+1+i\rho _{ji})}\,,  \label{gl}
\end{eqnarray}
where $\Theta$ is the Heaviside step function. In (\ref{FS4}), the sum
$\sum_{(p_{1},p_{2})}$ is over the two pairs $(p_{1},p_{2})$ and
$(p_{2},p_{1})$ solutions of the energy-momentum conservation laws: 
\begin{equation}
W=-\pi (\sin \,p_{1}+\sin \,p_{2})\,;\qquad \quad K=-p_{1}-p_{2}\,.
\label{conservation-laws}
\end{equation}
They read: 
\begin{equation}
(p_{1},p_{2})=\left( -K/2+\arccos \left( W/\left[ 2\pi \sin \left(
K/2\right) \right] \right) \,,\;-K/2-\arccos \left( W/\left[ 2\pi \sin
\left( K/2\right) \right] \right) \right) .
\label{solution-conservation-laws}
\end{equation}
Note that the solution in (\ref{solution-conservation-laws}) is allowed as
long as $W_{l}\leq W\leq W_{u}$ where $W_{u}$ is given in
(\ref{arguments-of-F}) and: 
\begin{equation}
W_{l}=\pi |\sin \,K|\,.  \label{Wl}
\end{equation}

The (analytic) behavior of the function $F$ in (\ref{FS4}) is discussed in 
\cite{ABS}. It is shown that the series $g_{\ell }$ is convergent. It is
also shown that $g_{\ell }$ stays finite when two $\rho _{i}$'s or more get
equal. Since the function $\exp \left( -h\right) $ goes to zero in these
regions \cite{KMB}, the integrand $F$ of $S_{4}$ is regular there.
Furthermore, it is shown that $F$ is exponentially convergent when one of
the $\rho _{i}$'s goes to infinity, which means the two integrals over $p_{3}
$ and $p_{4}$ in (\ref{S4})\ do not yield infinities. All these analytic
results help secure safe numerical manipulations.

\section{Behavior of exact four-spinon DSF}

From now on, we restrict ourselves to the interval $0\leq k\leq \pi $. All
forthcoming results can be carried to the other intervals of $k$ using the
symmetry relations (\ref{symmetries-of-DSF}). Also, we scale both $S_{4}$
and $S_{2}$ to appropriate units in order to display conveniently their
respective behaviors.

\subsection{Four-spinon continuum}

The first feature we discuss is the `four-spinon continuum', by analogy with
the two-spinon (or the spin-wave) continuum. It is the extent of the region
in the $(k,\omega )$-plane outside which $S_{4}$ is identically zero.
Remember that from (\ref{S2exact}), $S_{2}$ is confined to the region
$\omega _{2l}(k)\leq \omega \leq \omega _{2u}(k)$, where $\omega _{2l,u}(k)$
are the dCP boundaries given in (\ref{dCP}). From the condition $W_{l}\leq
W\leq W_{u}$ mentioned after (\ref{solution-conservation-laws}), we deduce
that in order for $S_{4}$ to be nonzero identically, we must have $\omega
_{4l}(k)\leq \omega \leq \omega _{4u}(k)$, where: 
\begin{eqnarray}
\omega _{4l}(k) &=&3\pi \sin (k/3)\quad \mathrm{for}\quad 0\leq k\leq \pi
/2\,;  \nonumber \\
\omega _{4l}(k) &=&3\pi \sin (k/3+2\pi /3)\quad \mathrm{for}\quad \pi /2\leq
k\leq \pi \,;  \nonumber \\
\omega _{4u}(k) &=&4\pi \cos (k/4)\quad \mathrm{for}\quad 0\leq k\leq \pi \,.
\label{dCP4}
\end{eqnarray}
We see that $\omega _{4l}(k)$ and $\omega _{4u}(k)$ are a sort of
four-spinon dCP boundaries for $S_{4}$. The two and four-spinon continua are
drawn in figure~1 below. We immediately notice that the four-spinon
continuum is not restricted to the region between the two-spinon dCP
branches, which means, a fortiori, that the full $S$ is also not confined to
the spin-wave continuum. This is a direct and explicit theoretical
confirmation of the `tail' of the dynamic structure function observed
outside the spin-wave continuum in finite-chain numerical calculations
\cite{Muller} and the phenomenology \cite{Exp}.
Due to the six and higher spinon
contributions, it is actually legitimate to expect the full DSF to tail even
further outside the four-spinon continuum, with arguably much smaller
values. For example, in the interval $0\leq k/\pi <0.51741$, there is a
narrow region between $\omega _{2u}$ and $\omega _{4l}$ inside which both
$S_{2}$ and $S_{4}$ are identically zero whereas the total $S$ may have (very
small) nonzero values, something that could eventually be checked in
finite-chain calculations. However, it is not possible to estimate exactly
the continua corresponding to the $S_{n>4}$ without manageable explicit
formulae for these. But what is already clear from our study is the fact
that indeed, the spin-wave continuum is not restrictive to the total dynamic
structure function.

\FRAME{fhFUX}{11.2028cm}{8.5624cm}{0pt}{\Qcb{Four-spinon (full) and
two-spinon (dashed) continua.}}{\Qlb{continuum}}{continuum.eps}
{\special{language "Scientific Word";type "GRAPHIC";display "FULL";valid_file
"F";width 11.2028cm;height 8.5624cm;depth 0pt;original-width
7.76in;original-height 11.1855in;cropleft "-0.0269";croptop
"0.6353";cropright "1";cropbottom "-0.0289";filename
'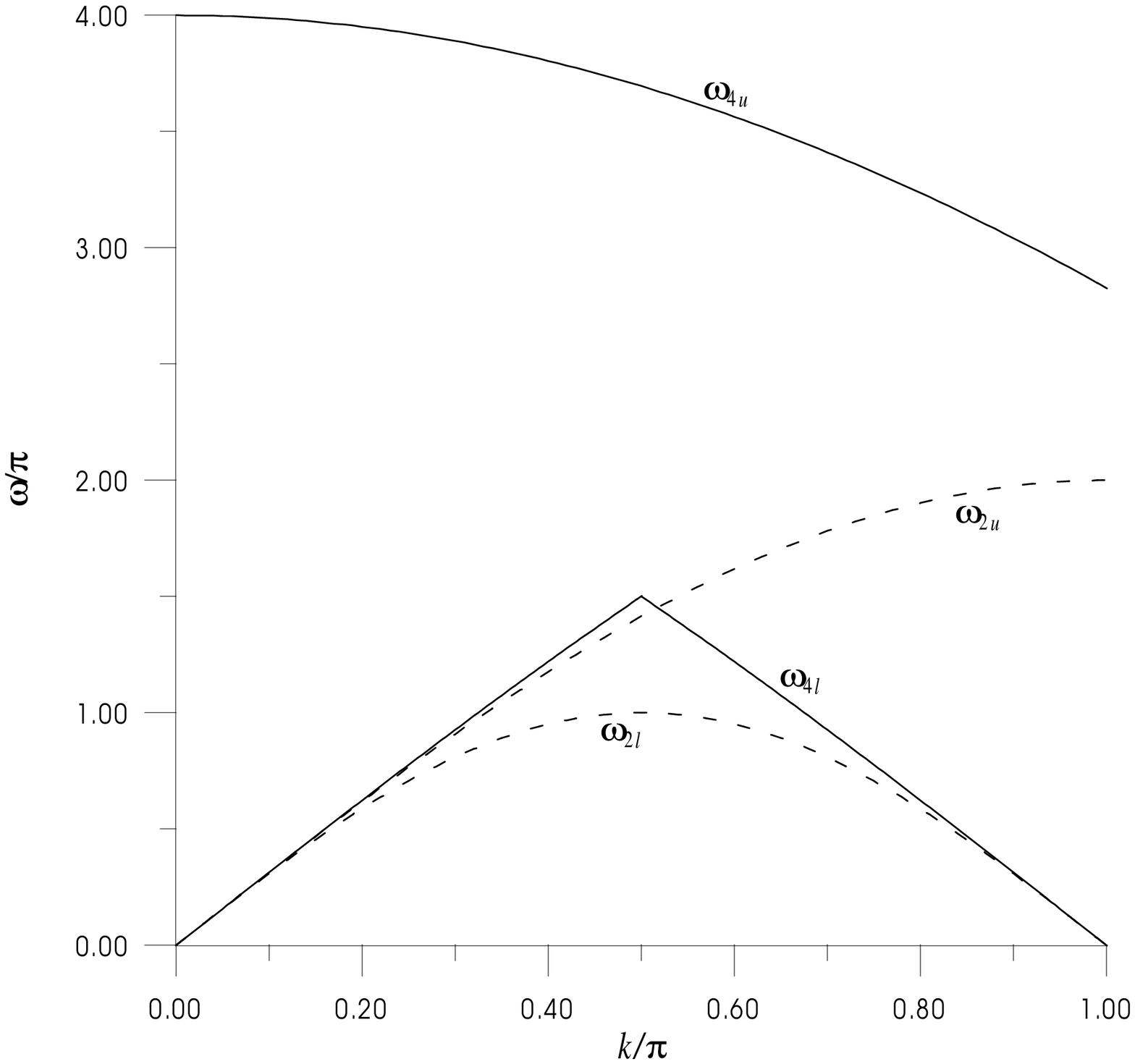';file-properties "XNPEU";}}

There are further general features we can also read from figure~1 without
having recourse to detailed calculations. For example, we see that inside
the interval $0\leq k/\pi <\allowbreak 0.51741$, the four-spinon continuum
lies entirely above the two-spinon continuum. This means that for this
interval, $S_{2}$ may be accepted as a good approximation for the total $S$
between the spin-wave boundaries and $S_{4}$ between $\omega _{4l}$ and
$\omega _{4u}$. For $0.51741\leq k/\pi \leq 1$ however and as $k$ increases,
there is increasing overlap between the two continua so that inside the
spin-wave continuum, we may expect $S$ to divert in\ perhaps a
non-negligible way from $S_{2}$.

\subsection{Behavior as a function of the energy transfer}

Next we describe the behavior of $S_{4}$ as a function of $\omega $ for
fixed values of $k$. Figure 2 displays the shapes of $S_{4}$ for
respectively $k/\pi =1/4,1/2,3/4$ and $1$. As an illustration, take for
example the case $k/\pi =1/4$. We see on figure 2 that $S_{4}$ is zero until 
$\omega /\pi \simeq 0.68$, which corresponds to the beginning of the
four-spinon continuum $\omega _{4l}(\pi /4)/\pi =3\sin (\pi /12)=
0.77646$. We obtain a slightly smaller value when reading directly from
figure 2 because of fitting. Note that it is important to check each time
consistency with the four-spinon continuum. This is because the lower and
upper branches $\omega _{4l,u}(k)$ are not imposed explicitly in the
integration algorithm of $S_{4}$ unlike the case of $S_{2}$ where the
corresponding expression (\ref{S2exact}) incorporates explicitly the
two-spinon dCP boundaries $\omega _{2l,u}(k)$. For $S_{4}$, only the
conditions $W_{l}\leq W\leq W_{u}$ noted just before (\ref{Wl}) are imposed.

\FRAME{fhFUX}{12.9118cm}{12.9447cm}{0pt}{\Qcb{(Scaled) $S_{4}$ as a function
of $\protect\omega $ for fixed $k$.}}{\Qlb{S4k}}{s4k.eps}{\special{language
"Scientific Word";type "GRAPHIC";display "USEDEF";valid_file "F";width
12.9118cm;height 12.9447cm;depth 0pt;original-width 7.76in;original-height
11.1855in;cropleft "0.0676";croptop "0.9116";cropright "0.9363";cropbottom
"0.0878";filename '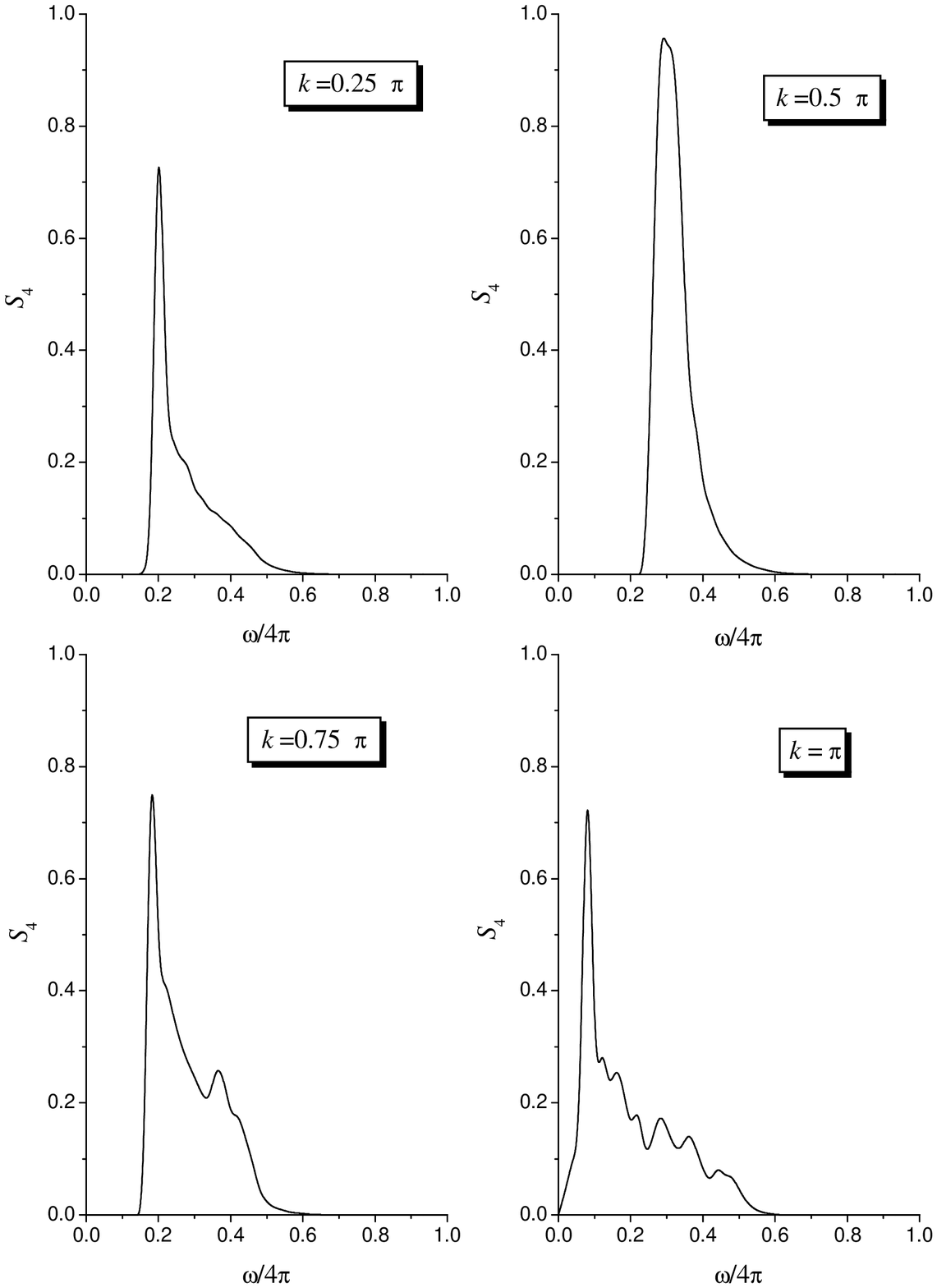';file-properties "XNPEU";}}

\FRAME{fhFUX}{12.7031cm}{12.9381cm}{0pt}{\Qcb{(Scaled) $S_{2}$ as a function
of $\protect\omega $ for fixed $k$.}}{\Qlb{s2k}}{s2k.eps}{\special{language
"Scientific Word";type "GRAPHIC";display "USEDEF";valid_file "F";width
12.7031cm;height 12.9381cm;depth 0pt;original-width 7.76in;original-height
11.1855in;cropleft "0.0726";croptop "0.9117";cropright "0.9271";cropbottom
"0.0882";filename '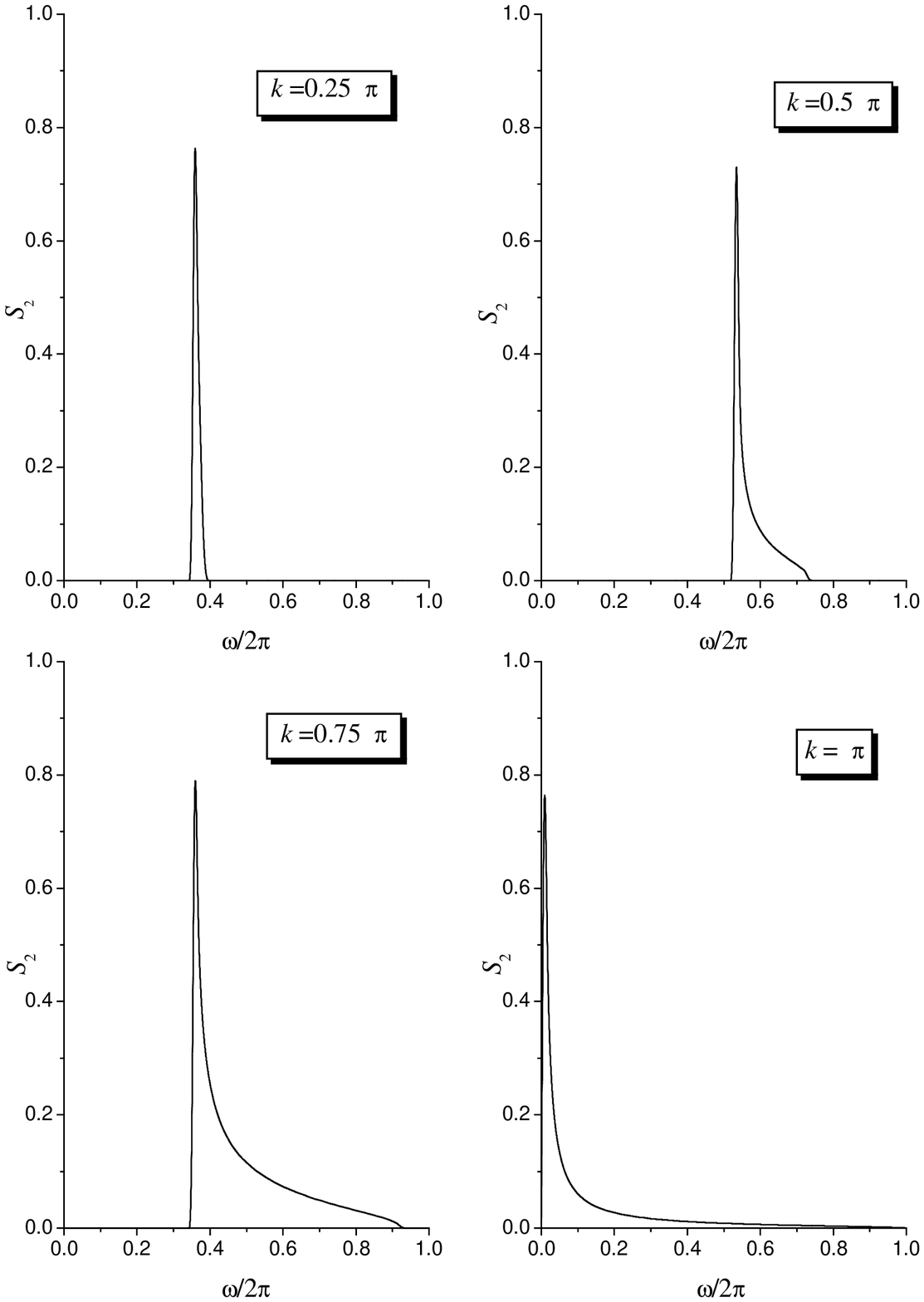';file-properties "XNPEU";}}

Starting from $\omega /4\pi \simeq 0.17$, $S_{4}$ jumps sharply from zero to
a maximum at $\omega /4\pi \simeq 0.21$ (read from figure 2). Then it
decreases with two apparent local minima until it becomes negligible at
roughly $\omega /4\pi \simeq 0.6$. The upper branch of the four-spinon
continuum at $k=\pi /4$ is $\omega _{4u}(\pi /4)/4\pi =\cos (\pi
/16)=\allowbreak 0.9808$. Hence, $S_{4}$ is practically negligible (but not
identically zero) for values of $\omega /4\pi $ between about $0.6$ and
$0.9808$. The description of $S_{4}$ as a function of $\omega$ for the other
values of $k$ can be carried along the same lines and each time, consistency
with the four-spinon continuum is checked. The shapes are all similar to one
another: steep increase from zero to a maximum value, followed by a
`wiggled' slower decrease to zero.

Furthermore, it is interesting to notice that for all values of $k$, the
overall shape of $S_{4}$, \textit{not} the detail, is roughly similar to the
one of $S_{2}$, represented as a function of $\omega$ for the same values
of $k$ in figure 3. For instance, for $k/\pi =1/4$, $S_{2}$ starts very
sharply at about $\omega /2\pi \simeq 0.35$, which corresponds to the start
of the two-spinon continuum at $\omega _{2l}(\pi /4)/2\pi = 0.35355$.
It reaches a maximum before decreasing more slowly towards zero.
The only difference worth mentioning is the decrease of $S_{4}$ after its
absolute maximum which presents local minima and maxima whereas the decrease
of $S_{2}$ is always `smooth'. This may originate from the `richer'
structure of the expression of $S_{4}$ with respect to that of $S_{2}$.

\subsection{Behavior as a function of the momentum transfer}

Last we describe the behavior of $S_{4}$ as a function of $k$ for fixed
values of $\omega$. In figure 4 are plotted the graphs of $S_{4}$ in terms
of $k$ for $\omega /\pi =1/4$,  $1/2$, $3/4$, $1$ respectively. Let us
describe for example the case $\omega /\pi =1/4$. We see on figure 4 that
$S_{4}$ is zero until we reach the value $k/\pi \simeq 0.88$. On the other
hand, the four-spinon continuum lies outside the interval $0.07967\leq k/\pi
\leq 0.92033$. In the region $0\leq k/\pi \leq 0.0796\,7$, figure 4 shows no
discernible finite values for $S_{4}$, only a very thin `trace' that would
be more visible with a better resolution. This means that $S_{4}$ is
negligible for those small values of $k$. For larger values of $\omega$
though, $S_{4}$ picks up clear finite values in the interval $0\leq k\leq
3\arcsin (\omega /3\pi )$; see the other graphs on figure 4. Those values
get larger as $\omega$ increases.

\FRAME{fhFUX}{12.9711cm}{12.9381cm}{0pt}{\Qcb{(Scaled) $S_{4}$ as a function
of $k$ for fixed $\protect\omega $.}}{\Qlb{S4w}}{s4w.eps}{\special{language
"Scientific Word";type "GRAPHIC";display "USEDEF";valid_file "F";width
12.9711cm;height 12.9381cm;depth 0pt;original-width 7.76in;original-height
11.1855in;cropleft "0.0636";croptop "0.9117";cropright "0.9363";cropbottom
"0.0882";filename '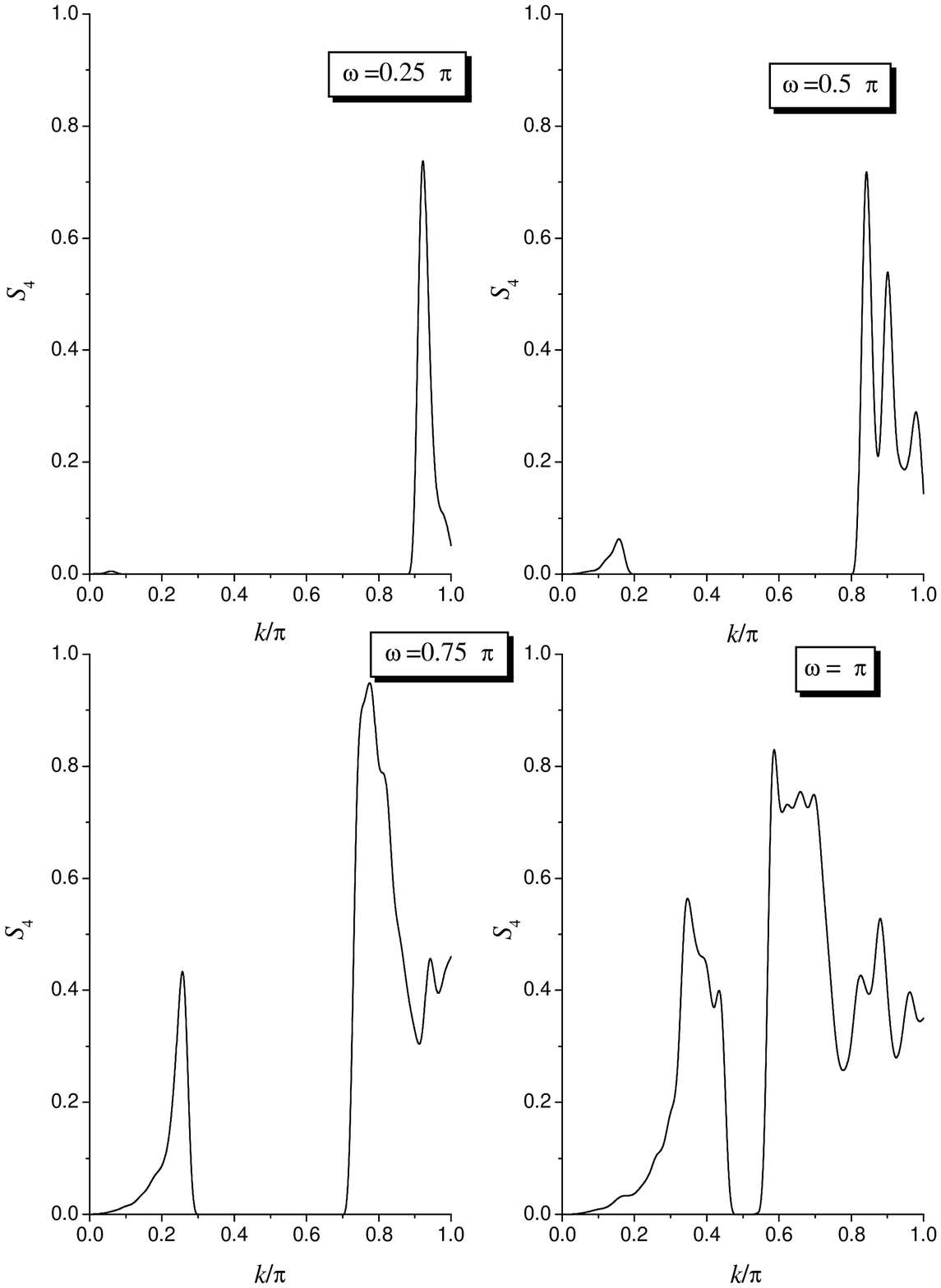';file-properties "XNPEU";}}

\FRAME{fhFUX}{13.9156cm}{12.9381cm}{0pt}{\Qcb{(Scaled) $S_{2}$ as a function
of $k$ for fixed $\protect\omega $.}}{\Qlb{S2w}}{s2w.eps}{\special{language
"Scientific Word";type "GRAPHIC";display "USEDEF";valid_file "F";width
13.9156cm;height 12.9381cm;depth 0pt;original-width 7.76in;original-height
11.1855in;cropleft "0.0317";croptop "0.9117";cropright "0.9682";cropbottom
"0.0882";filename '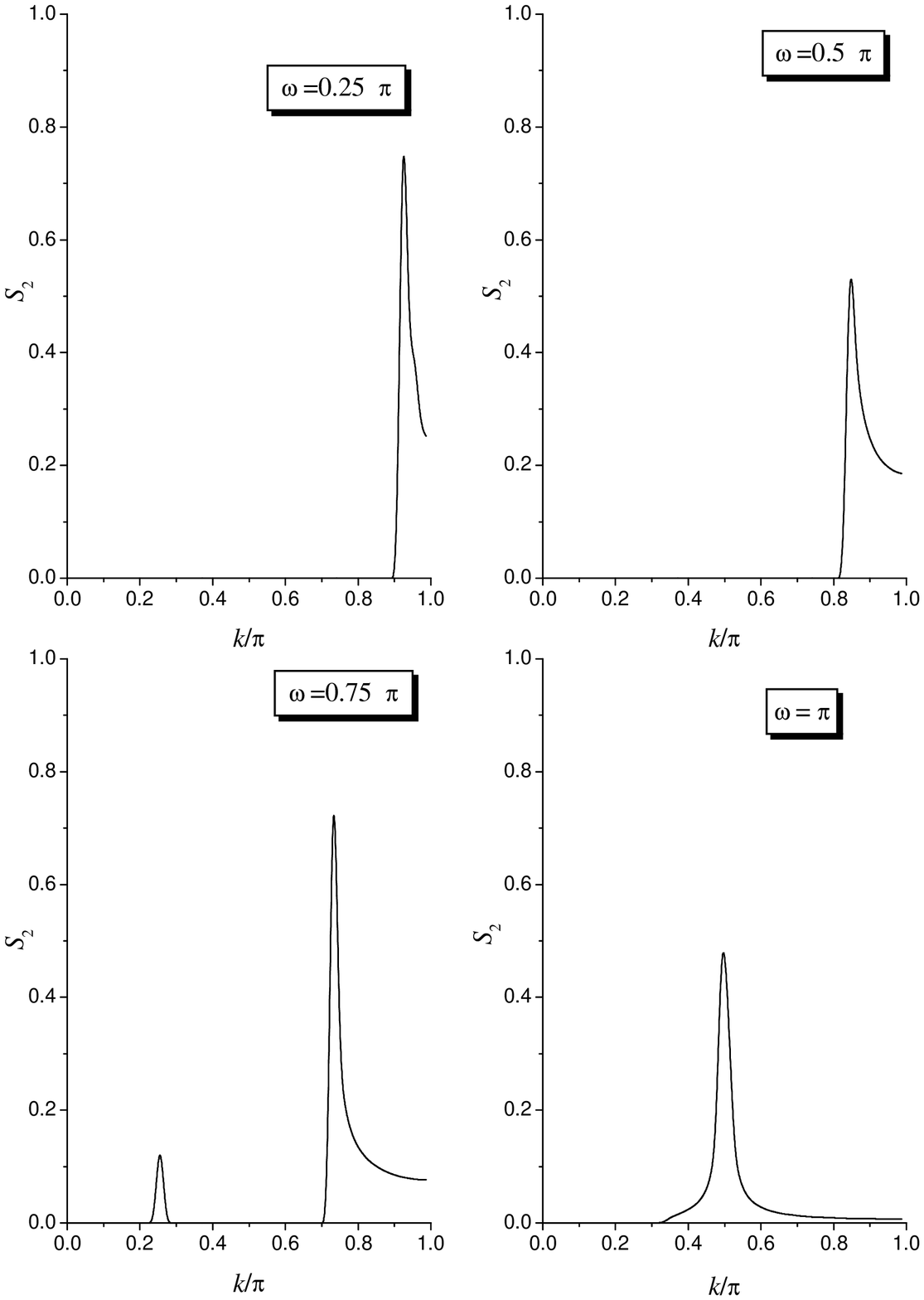';file-properties "XNPEU";}}

Returning to the case $\omega /\pi =1/4$, we see that $S_{4}$ starts from
zero at $k/\pi \simeq 0.88$ (slightly smaller than the exact value $0.92033$
because of fitting), rises sharply to a maximum and then decreases. The
behavior of $S_{4}$ for the other values of $\omega$ is also consistent
with the four-spinon continuum. Take for example the case $\omega /\pi =1/2$.
The four-spinon continuum indicates that $S_{4}$ is identically zero for
$0.1599\leq k/\pi \leq 0.84008$. We see
indeed small values for $S_{4}$ from $k/\pi =0$ up to a\ little before 0.2,
and $S_{4}$ rising again from zero a little after $k/\pi =0.8$ to a local
maximum, then to an absolute maximum before decreasing.

\FRAME{fhFUX}{14.8514cm}{6.4405cm}{0pt}{\Qcb{The integrand $F$ in (  \protect
\ref{S4}) as a function of $\left( p_{3},p_{4}\right) $ for $\left( k,%
\protect\omega \right) =\left( 0.52\protect\pi ,2\protect\pi \right) $.}}{%
\Qlb{F-1}}{f-1.eps}{\special{language "Scientific Word";type
"GRAPHIC";display "USEDEF";valid_file "F";width 14.8514cm;height
6.4405cm;depth 0pt;original-width 7.76in;original-height 11.1855in;cropleft
"0";croptop "0.6900";cropright "1";cropbottom "0.2822";filename
'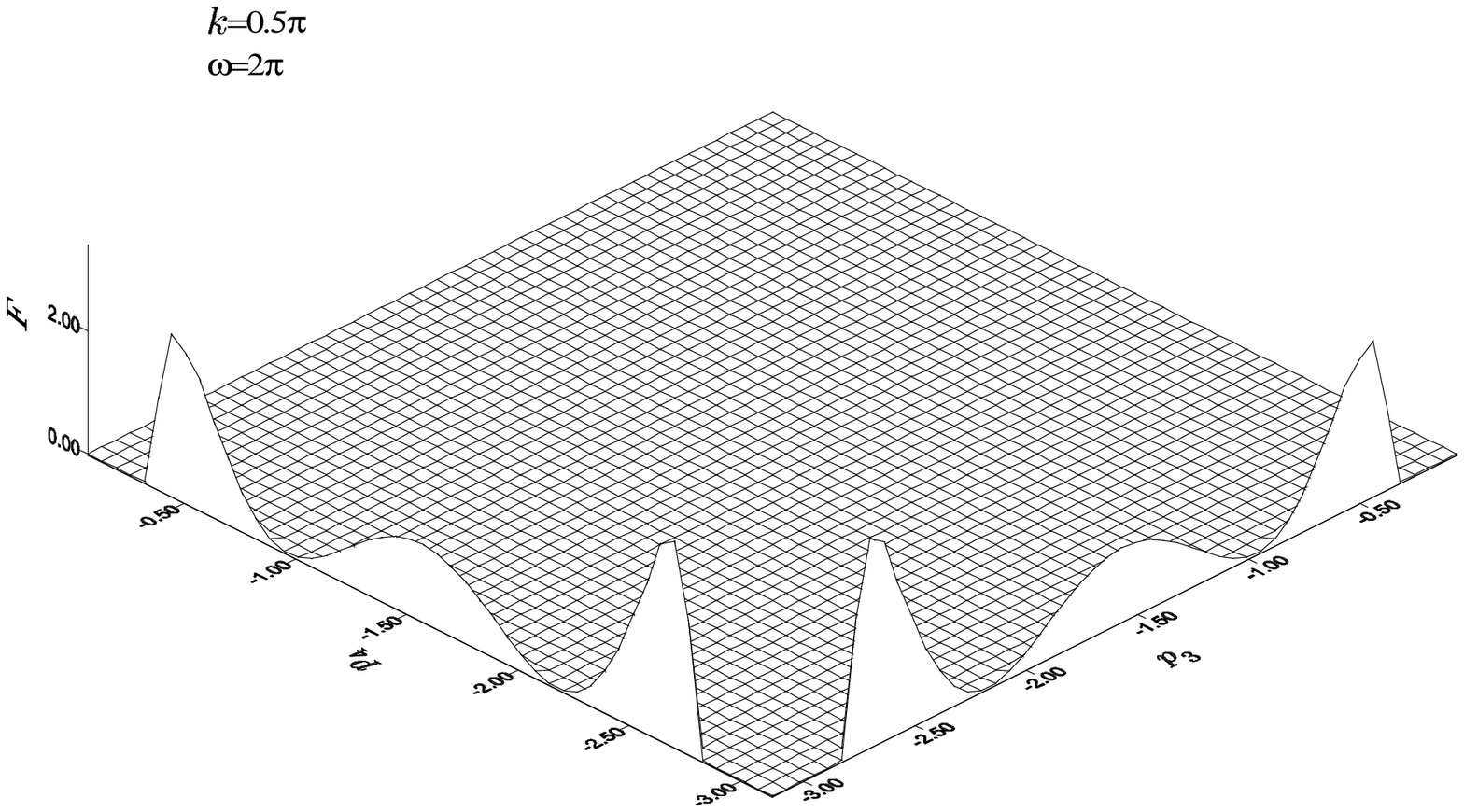';file-properties "XNPEU";}}

\FRAME{fhFUX}{14.8514cm}{6.4581cm}{0pt}{\Qcb{The integrand $F$ in ( \protect
\ref{S4}) as a function of $\left( p_{3},p_{4}\right) $ for $\left( k,%
\protect\omega \right) =\left( 0.5\protect\pi ,3\protect\pi \right) $.}}{%
\Qlb{F-2}}{f-2.eps}{\special{language "Scientific Word";type
"GRAPHIC";display "USEDEF";valid_file "F";width 14.8514cm;height
6.4581cm;depth 0pt;original-width 7.76in;original-height 11.1855in;cropleft
"0";croptop "0.6755";cropright "1";cropbottom "0.2667";filename
'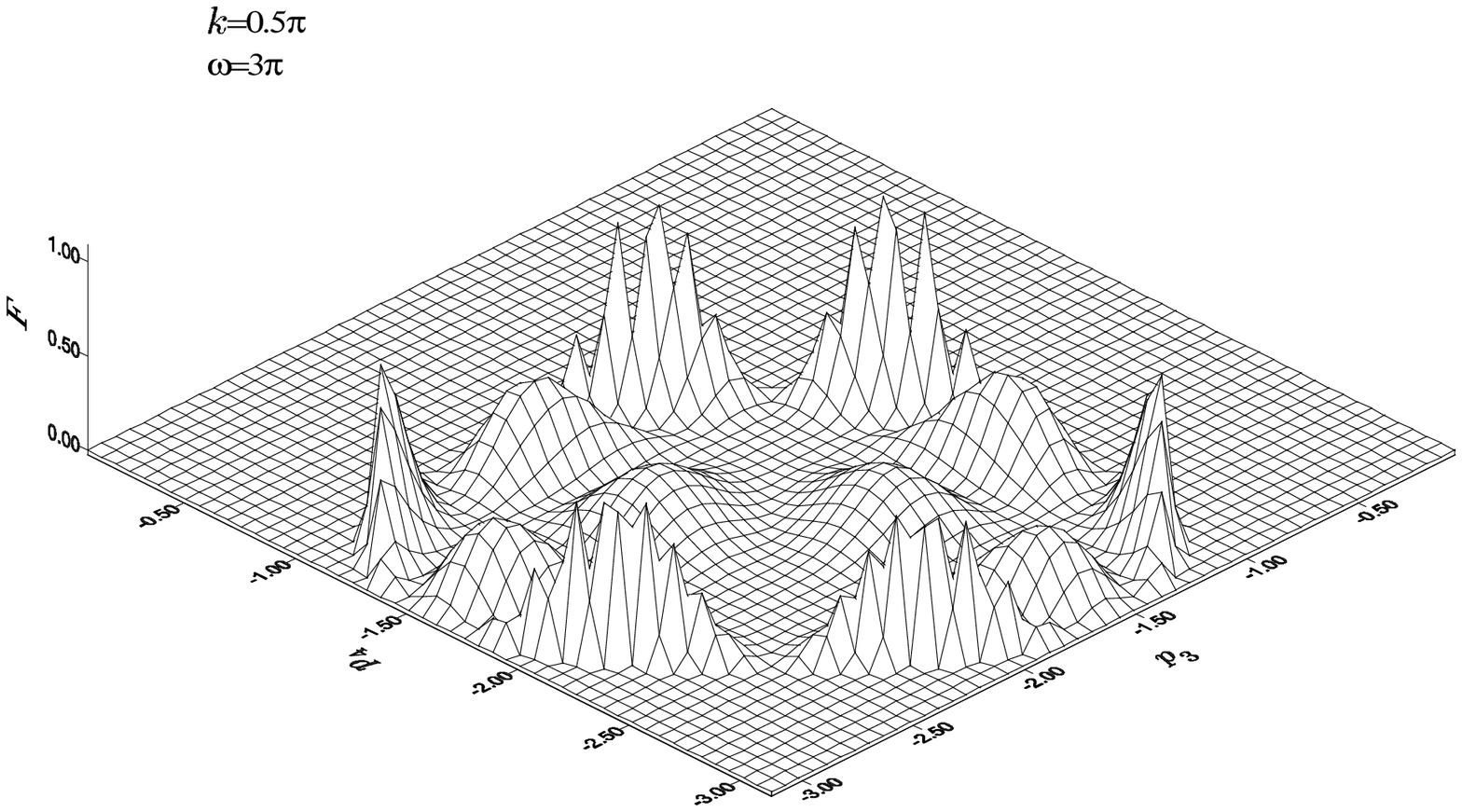';file-properties "XNPEU";}}

Here too $S_{2}$ has a similar overall behavior within its own (two-spinon)
continuum. Figure 5 displays the behavior of $S_{2}$ with respect to $k$ for
the same fixed values of $\omega $. But as $\omega$ increases, we notice
the richer structure of $S_{4}$ with respect to the corresponding one for
$S_{2}$. This is due to the more involved expression of $S_{4}$. In fact,
for larger values of $\omega$, the integrand $F$ in (\ref{S4}) is
nonvanishing in larger and larger areas in the $\left( p_{3},p_{4}\right)$
plane. For illustration, compare the behavior of $F$ shown in figure\ 6 for
$\left( k,\omega \right) =\left( 0.5\pi ,2\pi \right)$ with the one shown in
figure 7 for $\left( k,\omega \right) =\left( 0.5\pi ,3\pi \right)$. In
this regard, we recall that the quadrature-based algorithms written in
\cite{ABSS} were based on this observation and took advantage of the fact
that for fairly small values of $\omega$, the integrand $F$ was negligible in
large areas of the $\left( p_{3},p_{4}\right)$ plane. This is the reason
why those algorithms could not be carried to larger values of $\omega$, or
be able to describe efficiently a behavior as a function of the neutron
energy $\omega$ itself. As already mentioned, we have used here Monte Carlo
techniques and we can do better, but consistency\ with \cite{ABSS} is
realized, i.e., we have the same behavior of $S_{4}$ as a function of $k$
for the same small values of $\omega$.

\section{Conclusion}

In this work, we have described the behavior of the exact four-spinon
dynamic structure function $S_{4}$ in the antiferromagnetic isotropic
Heisenberg quantum spin chain at zero temperature as a function of the
neutron energy $\omega$ and momentum transfer $k$. We have also determined
the four spinon continuum, the region outside which $S_{4}$ is identically
zero. The discussion was carried in the form of a comparison with the
corresponding behavior of the exact two-spinon dynamic structure function
$S_{2}$, already known in the literature. Figures 8 and 9 summarize these two
behaviors where $S_{4}$ (8) and $S_{2}$ (9) are drawn in the $\left(
k,\omega \right) $ plane. Recall that we have scaled them down to one.

\FRAME{fhFUX}{14.3813cm}{7.3587cm}{0pt}{\Qcb{(Scaled) $S_{4}$ as a function
of $k$ and $\protect\omega $.}}{\Qlb{S4kw}}{s4kw.eps}{\special{language
"Scientific Word";type "GRAPHIC";display "USEDEF";valid_file "F";width
14.3813cm;height 7.3587cm;depth 0pt;original-width 7.76in;original-height
11.1855in;cropleft "0";croptop "0.6719";cropright "0.9681";cropbottom
"0.2773";filename '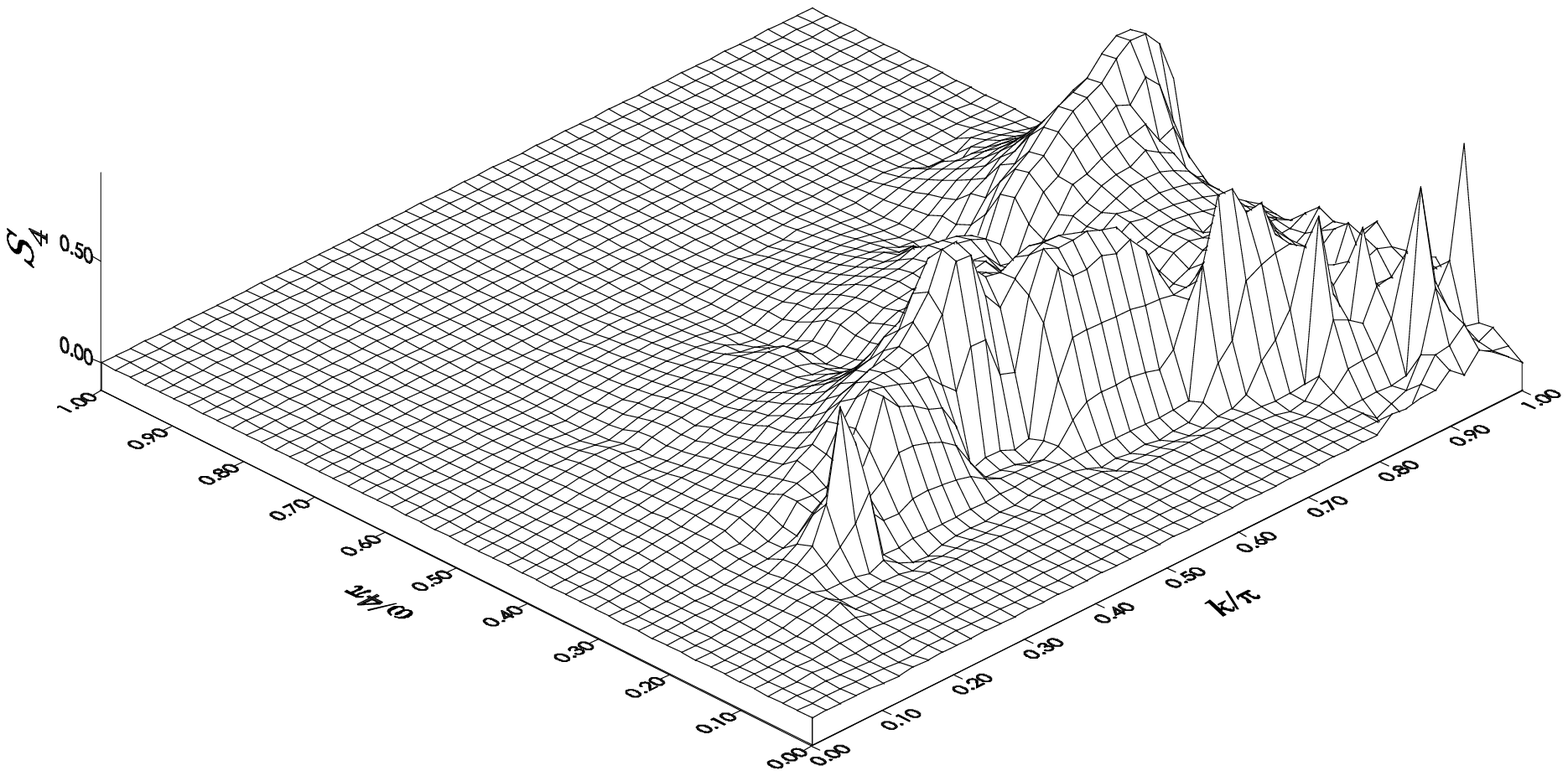';file-properties "XNPEU";}}

\FRAME{fhFUX}{14.8514cm}{6.5789cm}{0pt}{\Qcb{(Scaled) $S_{2}$ as a function
of $k$ and $\protect\omega $.}}{\Qlb{s2kw}}{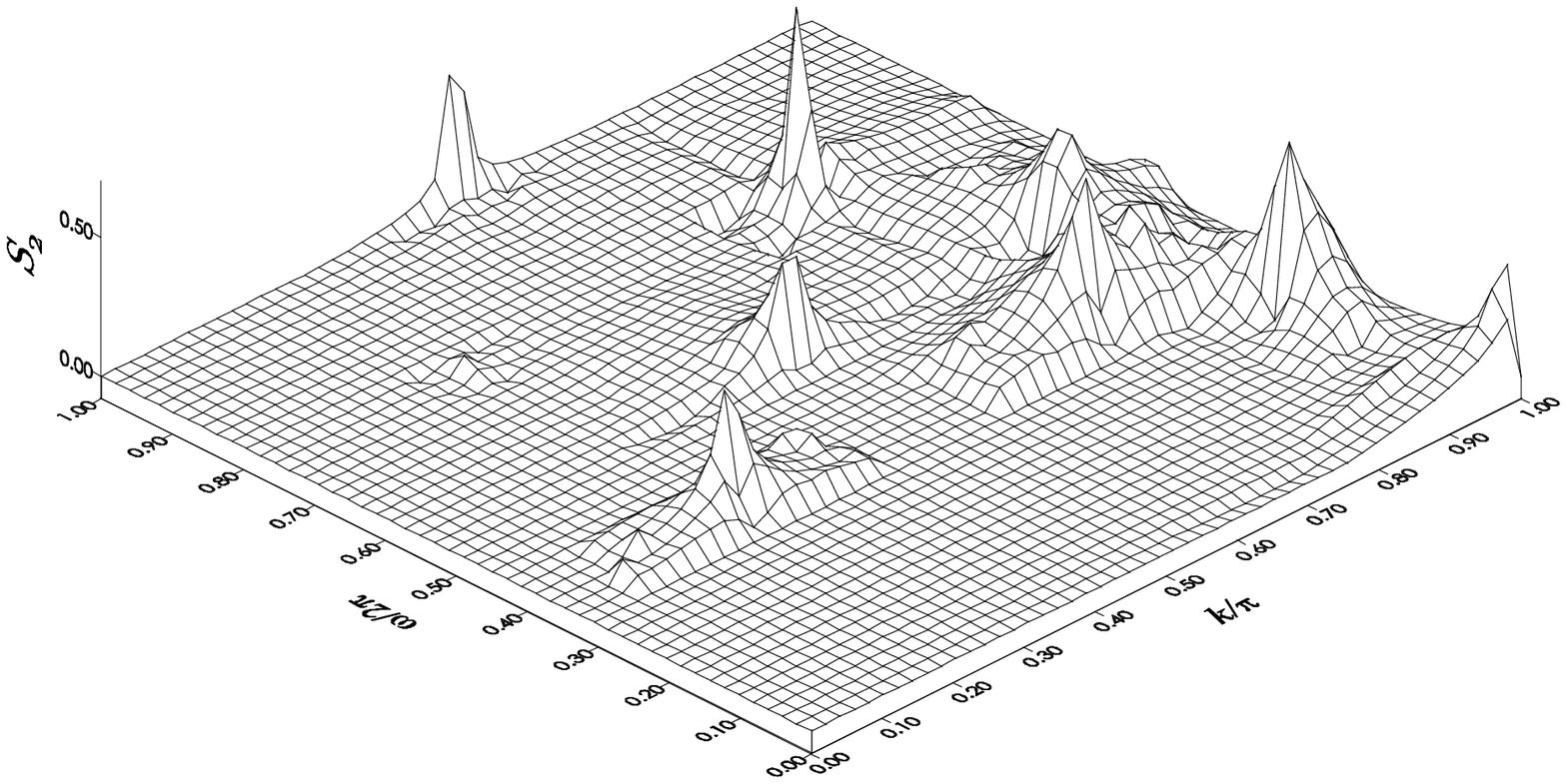}{\special{language
"Scientific Word";type "GRAPHIC";display "USEDEF";valid_file "F";width
14.8514cm;height 6.5789cm;depth 0pt;original-width 7.76in;original-height
11.1855in;cropleft "0";croptop "0.6466";cropright "1";cropbottom
"0.2942";filename 's2kw.eps';file-properties "XNPEU";}}

There are four directions in which one may wish to carry forward with this
work. The first is the anisotropic case. The model is exactly soluble and we
do have generic expressions for $S_{n}$ in the form of contour integrals in
the spectral parameters complex planes \cite{Boug}. The difficulty here is
that the integrands involve much more complicated functions which are
already present in $S_{2}$, and one should expect intricate complexities in
this more general case.

The second direction in which one may want to push forward is the situation
where there is an external magnetic field. There are finite-chain
calculations in this regard, \cite{Muller} and more recently the works
\cite{KM-KBM}. But one has to remember that the model in not exactly solvable in
this case. So one may want to try small perturbations around the zero-field
limit solution. The third direction is the finite-temperature case. Here too
there are finite-chain results and it is interesting to see those effects on 
$S_{2}$ and $S_{4}$. The fourth direction is to look into the situation of a
spin-one chain. The model is still exactly solvable and exploiting the
quantum group symmetry, compact expressions for the form factors are
available \cite{idzumi,bougourzi-weston,bougourzi-any-spin}.


\begin{thebibliography}{99}
\bibitem{hirakawa-kurogi} K. Hirakawa and Y. Kurogi, Prog.\ Theor. Phys. 
\textbf{S46} (1970)\ 147.

\bibitem{Exp} D.A.\ Tennant, R.A. Cowley, S.E.\ Nagler and A.M. Tsvelik,
Phys. Rev. \textbf{B}52 (1995) 13368; D.A. Tennant, S.E. Nagler, D. Weltz,
G. Shirane and K. Yamada, Phys. Rev. \textbf{B}52 (1995) 13381; D.A.
Tennant, T.G. Perring, R.A. Cowley and S.E.\ Nagler, Phys. Rev. Lett. 70
(1993) 4003; S.E. Nagler, D.A.\ Tennant, R.A. Cowley, T.G. Perring and S.K.
Satija, Phys. Rev. \textbf{B}44 (1991) 12361.

\bibitem{squires-lovesey} G.L. Squires, `\textit{Introduction to the theory
of thermal neutron scattering'}, Cambridge University Press, 1996; S.W.
Lovesey, `\textit{Theory of neutron scattering from condensed matter'},
Clarendon, Oxford, 1987.

\bibitem{PSD} Some (not \textit{all}) of the literature related to the
Heisenberg model comprises W. Heisenberg, Z. Phys. 49 (1928) 619; H. Bethe,
Z. Phys. 71 (1931) 205; L. Hulth\'{e}n, Arkiv. Mat. Astron. Fysik A 1126
(1938) 1; E.H. Lieb and D.C. Mattis, J. Math. Phys. 3 (1962) 749; J.~des
Cloizeaux and J.J. Pearson, Phys. Rev. 128 (1962) 2131; R.B. Griffiths,
Phys. Rev. 133 (1964) A 768; C.N. Yang and C.P. Yang, Phys. Rev. 150
(1966) 321; 150 (1966)\ 327; 151 (1966) 258; Th. Niemeijer, Physica 36
(1967) 377; E. Barouch, B.M. McCoy and D.B. Abraham, Phys. Rev. \textbf{A}4
(1971) 2331; M. Gaudin, Phys. Rev. Lett. 26 (1971) 1301; M. Takahashi, Prog.
Theor. Phys. 46 (1971)\ 401; L.A. Takhtajan and L.D. Faddeev, Russ. Math.
Surveys 34 (1979) 11; B.M. McCoy, J.H.H. Perk and R.E. Shrock, Nucl. Phys. 
\textbf{B}220 (1983) 35; O. Babelon, H.J. de Vega and C.M. Viallet, Nucl.
Phys. \textbf{B}220 (1983) 13; G. M\"{u}ller and R.E. Shrock, Phys. Rev. 
\textbf{B}29 (1984)\ 288; J.M.R. Roldan, B.M. McCoy and J.H.H. Perk, Physica
136A (1986)\ 255; V.E. Korepin, A.G.\ Izergin and N.M. Bogoliubov,
`\textit{The Quantum Inverse Scattering Method and Correlation Functions}',
Cambridge University Press, 1993; F.H.L. Essler, H. Frahm, A.G. Izergin and
V.E. Korepin, Comm. Math. Phys. 174 (1994) 191; V.E. Korepin, A.G. Izergin,
F.H.L. Essler and D. Uglov, Phys. Lett. \textbf{A}190 (1994) 182.

\bibitem{Baxter} R.J. Baxter, `\textit{Exactly Solved Models in Statistical
Mechanics}', Academic Press, 1982.

\bibitem{JimMi} M. Jimbo and T. Miwa, `\textit{Algebraic Analysis of
Solvable Lattice Models}', American Mathematical Society, 1994.

\bibitem{CorFun} O. Davies, O. Foda, M. Jimbo, T. Miwa and A. Nakayashiki,
Comm. Math. Phys. 151 (1993) 89; I.B. Frenkel and N.H.\ Jing, Proc. Natl.
Acad. Sci. 85 (1988) 9373; A. Abada, A.H.\ Bougourzi and M.A. El Gradechi,
Mod. Phys. Lett. A8 (1993)\ 715; A.H. Bougourzi, Nucl. Phys. \textbf{B}404
(1993)\ 457; A.H.\ Bougourzi, `\textit{Bosonization of quantum affine groups
and its application to the higher spin Heisenberg model'},
\texttt{q-alg/9706015}.

\bibitem{And} P.W.\ Anderson, Phys. Rev. 86 (1952) 694.

\bibitem{TMMC} M.T. Hutchings, G. Shirane, R.J. Birgeneau and S.L. Holt,
Phys. Rev. \textbf{B}5, (1972)\ 1999.

\bibitem{Muller} G. M\"uller, H. Thomas, H. Beck and J.C. Bonner, Phys. Rev. 
\textbf{B}24 (1981) 1429.

\bibitem{BCK} A.H. Bougourzi, M. Couture and M. Kacir, Phys. Rev.
\textbf{B}54 (1996) 12669.

\bibitem{KMB} M. Karbach, G. M\"{u}ller and A.H. Bougourzi,
`\textit{Two-spinon dynamic structure factor of the one-dimensional $S=1/2$
Heisenberg antiferromagnet}', \texttt{cond-mat/9606068}.

\bibitem{BKM} A.H. Bougourzi, M. Karbach and G. M\"{u}ller, `\textit{Exact
two-spinon dynamic structure factor of the one-dimensional} $s=1/2$
\textit{Heisenberg-Ising antiferromagnet}', \texttt{cond-mat/9712101}.

\bibitem{ABS} A. Abada, A.H. Bougourzi and B. Si-lakhal, Nucl. Phys.
\textbf{B}497 [FS] (1997) 733.

\bibitem{Boug} A.H. Bougourzi, Mod. Phys. Lett \textbf{B}10 (1996) 1237.

\bibitem{ABSS} A. Abada, A.H. Bougourzi, B. Si-Lakhal and S. Seba,
`\textit{Four-Spinon Dynamical Correlation Function
in Isotropic Heisenberg Model'}, \texttt{cond-mat/9802271}, unpublished.

\bibitem{Fadeev-Takhtajan} L.D. Faddeev and L.A. Takhtajan, J. Soviet. Math.
24 (1984) 241.

\bibitem{smirnov} F.A. Smirnov, `\textit{Form Factors in Completely
Integrable Models of Quantum Field Theory}' World Scientific, Singapore,
1992.

\bibitem{KM-KBM} M. Karbach, D. Biegel and G. M\"{u}ller,
`\textit{Quasiparticles governing the
zero-temperature dynamics of the 1D spin-1/2
Heisenberg antiferromagnet in a magnetic field}', \texttt{cond-mat/0205142};
M.\ Karbach and G. M\"{u}ller, `\textit{Line shape predictions via Bethe
ansatz for the one dimensional spin-1/2 Heisenberg antiferromagnet in a
magnetic field}', \texttt{cond-mat/0005174}.

\bibitem{idzumi} M. Idzumi, Int. J. Mod. Phys. \textbf{A}9 (1994) 4449;
`\textit{Correlation functions of the spin 1 analog of the XXZ model}', 
\texttt{hep-th/9307129}.

\bibitem{bougourzi-weston} A.H. Bougourzi and R.A. Weston, Nucl. Phys. 
\textbf{B}417 (1994) 439.

\bibitem{bougourzi-any-spin} A.H. Bougourzi, `\textit{Bosonization of
quantum affine groups and its application to higher spin Heisenberg model}', 
\texttt{q-alg/9706015}.
\end{thebibliography}
\end{document}